\documentclass[draft,showpacs,showkeys,eqsecnum,nofootinbib]{revtex4}
\renewcommand{\theequation}{\arabic{equation}}
\def\beq{\begin{equation}}
\def\eeq{\end{equation}}
\def\bea{\begin{eqnarray}}
\def\eea{\end{eqnarray}}
\def\nn{\nonumber}

\def\pa{\partial}

\begin{document}
\title{BRST invariance and de Rham-type cohomology of 't Hooft-Polyakov monopole}
\author{Soon-Tae Hong}
\email{soonhong@ewha.ac.kr} \affiliation{Department of Science
Education and Research Institute for Basic Sciences,\\ Ewha Womans
University, Seoul 120-750 Korea}
\date{\today}%
\begin{abstract}
We exploit the 't Hooft-Polyakov monopole to define closed algebra
of the quantum field operators and the BRST charge $Q_{BRST}$.
In the first-class configuration of the Dirac
quantization, by including the $Q_{BRST}$-exact gauge fixing term
and the Faddeev-Popov ghost term, we find the BRST invariant
Hamiltonian to investigate the de Rham-type cohomology group
structure for the monopole system. The Bogomol'nyi bound is also
discussed in terms of the first-class topological charge defined
on the extended internal 2-sphere.
\end{abstract}
 \maketitle

\section{Introduction}

The Becchi-Rouet-Stora-Tyutin (BRST)~\cite{brst} symmetries have
been considerably studied in the constrained physical systems on
which many interesting physics phenomena in Nature are described.
It is well known that the solitons and monopoles~\cite{soliton}
are subject to the second-class constraints which can be
rigorously treated in the Dirac Hamiltonian quantization
scheme~\cite{dirac64}. Despite all the past successes~\cite{hong02pr}
of the quantization of the constrained
physical systems through the uses of the first-class configuration
of the Dirac Hamiltonian formalism and its corresponding BRST
mechanism, there still does not exist a comprehensive
understanding of the hidden geometry involved in the BRST symmetry
invariance of the constrained systems.

Since the Dirac monopole string~\cite{dirac31} was proposed to
quantize the electric charges of the particles of matter, there
has been lots of progress in understanding the properties of the
magnetic monopole systems. It was shown in the Wu-Yang
monopole theory~\cite{wu76} that the interaction of the Dirac
monopole with the electromagnetic field removes the string line
singularity inherent in the monopole. Recently the supersymmetric
aspects of a charged particle were investigated in the background
of these monopoles~\cite{hong05}. Moreover, it was found that the
charged particle in the Dirac monopole background possesses
the geometric features associated with cocyles~\cite{grossman85}.
For the unconstrained gauged Yang-Mills theory, the BRST symmetry
was also analyzed in terms of the cocycles and chiral anomaly~\cite{bau86}.

In this paper, we will introduce the 't Hooft-Polyakov
monopole~\cite{thooft74,polyakov74} to yield its BRST charge, de
Rham-type cohomology and closed algebra of the quantum field operators.
To do this, we will find the first-class Hamiltonian of the monopole,
since the 't Hooft-Polyakov monopole is classified as the second-class
system in the Dirac quantization formalism. We will then define the
monopole charge in the U(1) subgroup of the SU(2) gauge group in the
first-class configuration to investigate the Bogomol'nyi bound on the
extended internal 2-sphere. We will next obtain the explicit form of
the BRST invariant Hamiltonian and discuss the geometric aspects of the
corresponding de Rham-type cohomology.

In Section II, we will consider the 't Hooft-Polyakov monopole
action to construct the closed algebra of its quantum operators
and the first-class Hamiltonian. In Section III, in the
first-class configuration, we will study the monopole charge
defined in the U(1) subgroup of the SU(2) gauge group and the U(1)
gauge invariant electromagnetic fields, and discuss the
Bogomol'nyi bound of the first-class static energy of the 't
Hooft-Polyakov monopole. By including the ghost fields, we will
then construct the BRST invariant 't Hooft-Polyakov monopole
Hamiltonian and define the de Rham-type cohomology group. In
Section IV, we will summarize our results with some comments, and
in Appendix A we will list the first-class canonical variables.

\section{Closed algebra of quantum operators and first-class Hamiltonian}
\setcounter{equation}{0}
\renewcommand{\theequation}{\arabic{section}.\arabic{equation}}

In this section, in the Dirac quantization formalism, we will construct the
closed algebra of the quantum operators and the first-class Hamiltonian of the
't Hooft-Polyakov monopole whose Lagrangian is given by~\cite{thooft74,polyakov74},
\begin{equation}
L_{0}=\int
d^{3}x\left[-\frac{1}{4}G_{\mu\nu}^{a}G^{a\mu\nu}
+\frac{1}{2}D_{\mu}\phi^{a}D^{\mu}\phi^{a}-\frac{1}{4}\lambda(\phi^{a}\phi^{a}-F^{2})^{2}\right],
\label{lag}
\end{equation}
where $A_{\mu}^{a}$ ($a$=1, 2, 3) and $\phi^{a}$ are the SU(2)
non-abelian gauge fields and the real scalar Higgs fields,
respectively. The field strengths of $A_{\mu}^{a}$ and the
covariant derivatives of $\phi^{a}$ are defined as \bea
G_{\mu\nu}^{a}&=&\pa_{\mu}A_{\nu}^{a}-\pa_{\nu}A_{\mu}^{a}+g\epsilon^{abc}A_{\mu}^{b}A_{\nu}^{c},\nn\\
D_{\mu}\phi^{a}&=&\pa_{\mu}\phi^{a}+g\epsilon^{abc}A_{\mu}^{b}\phi^{c}.
\eea Using the Lagrangian (\ref{lag}), we readily obtain the
equations of motion for the fields ($\phi^{a}$,
$A_{\mu}^{a}$)~\cite{thooft74,polyakov74} \bea
D_{\mu}G^{a\mu\nu}-g\epsilon^{abc}(D^{\nu}\phi^{b})\phi^{c}&=&0,\nn\\
D_{\mu}D^{\mu}\phi^{a}+\lambda(\phi^{a}\phi^{a}-F^{2})\phi^{a}&=&0.\label{eoms}
\eea The canonical momenta ($\pi^{a}$, $\pi^{a}_{\mu}$) conjugate
to ($\phi^{a}$, $A_{\mu}^{a}$) are given by
\bea \pi^{a}&=&D_{0}\phi^{a},\label{momenta1}\\
\pi^{a}_{i}&=&G_{0i}^{a},\label{momenta2}\\
\pi^{a}_{0}&=&0. \label{momenta3} \eea  We also obtain the
canonical momentum $\pi_{\lambda}$ conjugate to the real scalar
multiplier field $\lambda$ given by \beq
\pi_{\lambda}=0.\label{momentum}\eeq

Exploiting the above momenta and the Legendre transformation, we
obtain the Hamiltonian \beq H=\int
d^{3}x\left[\frac{1}{2}\pi^{a}\pi^{a}+\frac{1}{2}\pi_{i}^{a}\pi_{i}^{a}
+\frac{1}{2}D_{i}\phi^{a}D_{i}\phi^{a}+\frac{1}{4}G_{ij}^{a}G_{ij}^{a}+\frac{1}{4}
\lambda(\phi^{a}\phi^{a}-F^{2})^{2}\right], \label{canH} \eeq
where we have used the gauge condition
$A_{0}^{a}=0$~\cite{thooft74,polyakov74}. The canonical variables
are subject to the non-vanishing Poisson brackets \bea
\{\phi^{a}(x),\pi^{b}(y)\}&=&\delta^{ab}\delta^{3}(x-y),\nn\\
\{A_{\mu}^{a}(x),\pi_{\nu}^{b}(y)\}&=&\delta^{ab}\delta_{\mu\nu}\delta^{3}(x-y),\nn\\
\{\lambda(x),\pi_{\lambda}(y)\}&=&\delta^{3}(x-y). \eea By
implementing the Dirac quantization scheme~\cite{dirac64}, we find
that our Hamiltonian system is subject to the following
second-class constraints \bea
\Omega_{1}&=&\phi^{a}\phi^{a}-F^{2}\approx 0,\label{const21}\\
\Omega_{2}&=&\phi^{a}\pi^{a}\approx 0. \label{const22} \eea Here
one notes that, in fact, the time evolution of the identity
(\ref{momentum}) yields the constraint (\ref{const21}). Moreover,
the identities (\ref{momenta3}) and (\ref{momentum}) are easily
shown to be the trivial first-class constraints decoupled from our
system of interest. With $\epsilon^{12}=-\epsilon^{21}=1$ this
second-class constraint algebra is given by
\beq\Delta_{kk^{\prime}}(x,y)=\{\Omega_{k}(x),\Omega_{k^{\prime}}(y)\}
=\epsilon^{kk^{\prime}}\phi^{a}\phi^{a}\delta^{3}(x-y).
\label{delta} \eeq

Next, we consider the Dirac brackets defined as \beq
\{A(x),B(y)\}_{D}=\{A(x),B(y)\} -\int
d^{3}z~d^{3}z^{\prime}\{A(x),\Omega_{k}(z)\}\Delta_{kk^{\prime}}^{-1}(z,z^{\prime})
\{\Omega_{k^{\prime}}(z^{\prime}),B(y)\},\label{diracb}\eeq where
$\Delta_{kk^{\prime}}^{-1}$ is the inverse of
$\Delta_{kk^{\prime}}$ in (\ref{delta}).  Using the definition
(\ref{diracb}) and performing the canonical quantization scheme
$i\{A,B\}_{D}\rightarrow [A_{op},B_{op}]$ we find the
non-vanishing quantum operator commutators of the variables \bea
{[ \phi^{a}(x),\pi^{b}(y)]}&=&i\left(\delta^{ab}
-\frac{\phi^{a}\phi^{b}}{\phi^{c}\phi^{c}}\right)\delta^{3}(x-y),\nn\\
{[\pi^{a}(x),\pi^{b}(y)]}&=&\frac{i}{\phi^{c}\phi^{c}}
\left(\phi^{b}\pi^{a}-\phi^{a}\pi^{b}\right)\delta^{3}(x-y),\nn\\
{[A_{\mu}^{a}(x),\pi_{\nu}^{b}(y)]}&=&i\delta^{ab}\delta_{\mu\nu}\delta^{3}(x-y),\nn\\
{[\lambda(x),\pi_{\lambda}(y)]}&=&i\delta^{3}(x-y),\eea where the
canonical quantum operators are given by  \bea
\pi^{a}&=&-i\left(\delta^{ab}-\frac{\phi^{a}\phi^{b}}{\phi^{c}\phi^{c}}\right)\pa_{\phi^{b}},\nn\\
\pi_{\mu}^{a}&=&-i\pa_{{A}_{\mu}^{a}},~~~\pi_{\lambda}=-i\pa_{\lambda}.\eea
We also find the closed algebra \bea
{[S^{a},S^{b}]}&=&\epsilon^{abc}S^{c},\nn\\
{[S^{a},T^{b}]}&=&\epsilon^{abc}T^{c},\nn\\
{[T^{a},T^{b}]}&=&0,\eea where \bea S^{a}&=&\int d^{3}x~
i\epsilon^{abc}\pi^{b}\phi^{c},\nn\\
T^{a}&=&\int d^{3}x~i\phi^{a}.\eea


Following the Hamiltonian quantization scheme for constrained
systems~\cite{dirac64,faddeev86,bft,niemi88,hong02pr}, we proceed
to convert the second-class constraints $\Omega_i=0$ $(i=1, 2)$
into the first-class ones. For this we introduce two canonically
conjugate St\"uckelberg fields $(\theta, \pi_{\theta})$ with
Poisson bracket \beq \{\theta(x),
\pi_{\theta}(y)\}=\delta^{3}(x-y). \label{phii} \eeq The strongly
involutive first-class constraints $\tilde{\Omega}_{i}$ are
constructed as a power series of the St\"uckelberg fields to yield
\begin{eqnarray}
\tilde{\Omega}_{1}&=&\Omega_{1}+2\theta,  \nonumber \\
\tilde{\Omega}_{2}&=&\Omega_{2}-\phi^{a}\phi^{a}\pi_{\theta},
\label{1stconst}
\end{eqnarray}
and their commutator is given by
\beq
\{\tilde{\Omega}_{i},\tilde{\Omega}_{j}\}=0.\label{firstomega}
\eeq In general, following the Dirac quantization scheme, we can
construct the first-class constraints satisfying the Lie algebra
\beq
\{\tilde{\Omega}_{i},\tilde{\Omega}_{j}\}=C_{ij}^{~~k}~\tilde{\Omega}_{k}.
\label{cijk} \eeq Since the first-class constraints are strongly
zero to yield $\{\tilde{\Omega}_{i},\tilde{\Omega}_{j}\}|{\rm
phy}\rangle=0$ from (\ref{cijk}), one does not have any
difficulties in construction of the quantum commutators and in
quantization of the given monopole system. In that sense, one has
degrees of freedom in taking a set of the first-class constraints.
For instance, the first-class constraints $\tilde{\Omega}_{i}$ in
(\ref{1stconst}) are a specific choice with $C_{ij}^{~~k}=0$. In
fact, the sets of the first-class constraints form an equivalent
family governed by the SO(2) group~\cite{hong01}.

Next, after some tedious algebra, we construct the first-class
Hamiltonian of (\ref{canH}) in terms of the original fields \bea
\tilde{H}&=&\int
d^{3}x\left[\frac{1}{2}\left(\pi^{a}-\phi^{a}\pi_{\theta}\right)
\left(\pi^{a}-\phi^{a}\pi_{\theta}\right)\frac{\phi^{c}\phi^{c}}{\phi^{c}\phi^{c}+2\theta}
+\frac{1}{2}\pi_{i}^{a}\pi_{i}^{a}+\frac{1}{2}D_{i}\phi^{a}D_{i}\phi^{a}\frac{\phi^{c}\phi^{c}
+2\theta}{\phi^{c}\phi^{c}}+\frac{1}{4}G_{ij}^{a}G_{ij}^{a}\right.\nonumber\\
&&\left.+\frac{1}{4}\lambda(\phi^{a}\phi^{a}-F^{2}+2\theta)^{2}\right].
\label{hct} \eea We note that this Hamiltonian is strongly
involutive with the first-class constraints, \beq
\{\tilde{\Omega}_{i},\tilde{H}\}=0. \eeq When we consider the time
evolution of the constraint algebra, as determined by computing
the Poisson brackets of the constraints with the Hamiltonian
(\ref{hct}), we readily see from the Poisson bracket $\{
\tilde{\Omega}_{1}, \tilde H \}=0$ that we need to improve the
Hamiltonian into the following, equivalent first-class
Hamiltonian, \beq \tilde{H}^{\prime}=\tilde{H}+\int
d^{3}x~\pi_{\theta}\tilde{\Omega}_{2}. \label{hctp} \eeq In fact,
this improved Hamiltonian generates the constraint algebra
\begin{eqnarray}
\{\tilde{\Omega}_{1},\tilde{H}^{\prime} \}&=&2\tilde{\Omega}_{2},
\nonumber\\
\{\tilde{\Omega}_{2},\tilde{H}^{\prime}\}&=&0.
\end{eqnarray}
Since the Hamiltonians $\tilde{H}$ and $\tilde{H}^{\prime}$ only
differ by a term which vanishes on the constraint surface, they
lead to an equivalent dynamics on the constraint surface. Finally,
the first-class canonical variables are explicitly constructed and
listed in Appendix A.

\section{Monopole charge, BRST symmetry and de Rham-type cohomology}
\setcounter{equation}{0}
\renewcommand{\theequation}{\arabic{section}.\arabic{equation}}

In this section, we will investigate the de Rham-type cohomology
group structure for the 't Hooft-Polyakov monopole system, after
constructing its first-class monopole charge and BRST charge.  To
do this, we first revisit the original 't Hooft-Polyakov monopole
Lagrangian in (\ref{lag}) to consider the monopole charge which is
defined in the U(1) subgroup of the SU(2) gauge group. The U(1)
gauge invariant electromagnetic fields $F_{\mu\nu}$ are defined
as~\cite{thooft74,polyakov74}\beq
F_{\mu\nu}=\bar{\phi}^{a}G^{a}_{\mu\nu}-\frac{1}{g}\epsilon^{abc}\bar{\phi}^{a}
D_{\mu}\bar{\phi}^{b}D_{\nu}\bar{\phi}^{c},\label{fmunu}\eeq  and
the topological current $k^{\mu}$ is also defined
as~\cite{arafune75} \beq
k^{\mu}=-\frac{1}{8\pi}\epsilon^{\mu\nu\rho\sigma}\epsilon^{abc}\pa_{\nu}\bar{\phi}^{a}
\pa_{\rho}\bar{\phi}^{b}\pa_{\sigma}\bar{\phi}^{c},\label{topcurr}\eeq
where the rescaled real scalar Higgs fields are given by\beq
\bar{\phi}^{a}=\frac{\phi^{a}}{(\phi^{c}\phi^{c})^{1/2}}.\eeq
Exploiting the conformal map condition \beq
\bar{\phi}^{a}D_{\mu}\bar{\phi}^{a}=\bar{\phi}^{a}\pa_{\mu}\bar{\phi}^{a}=0,\eeq
one readily checks that the dual equations of motion for the
electromagnetic fields $F_{\mu\nu}$ in (\ref{fmunu}) yield \beq
\frac{1}{2}\epsilon^{\mu\nu\rho\sigma}\pa_{\nu}F_{\rho\sigma}=\frac{4\pi}{g}k^{\mu},\label{dual}\eeq
from which the magnetic monopole charge $m$ is given by \beq
m=\frac{1}{g}\int d^{3}x~k^{0}=\frac{1}{g}Q_{top}. \eeq Here
$Q_{top}$ is the topological charge to be discussed later.

Next, we return to the first-class physical system described in
the previous section.  In this configuration, the first-class
topological current is given by \beq
\tilde{k}^{\mu}=-\frac{1}{8\pi}\epsilon^{\mu\nu\rho\sigma}\epsilon^{abc}\pa_{\nu}\bar{\phi}^{a}
\pa_{\rho}\bar{\phi}^{b}\pa_{\sigma}\bar{\phi}^{c}\left(\frac{\phi^{c}\phi^{c}
+2\theta}{\phi^{c}\phi^{c}}\right)^{3/2}.\label{topcurr2}\eeq Here
we have used the first-class rescaled fields
$\tilde{\bar{\phi}}^{a}$ defined as \beq
\tilde{\bar{\phi}}^{a}=\bar{\phi}^{a}\left(\frac{\phi^{c}\phi^{c}
+2\theta}{\phi^{c}\phi^{c}}\right)^{1/2},\eeq which satisfies \beq
\tilde{\bar{\phi}}^{a}\tilde{\bar{\phi}}^{a}-1=0.\label{constbart}\eeq
Exploiting the antisymmetric property of
$\epsilon^{\mu\nu\rho\sigma}$ in (\ref{topcurr}) and
(\ref{topcurr2}), one readily check the following divergence \beq
\pa_{\mu}\tilde{k}^{\mu}=0,\eeq as in the second-class case:
$\pa_{\mu}k^{\mu}=0$.  Moreover, the first-class magnetic monopole
charge $\tilde{m}$ is given by \beq
\tilde{m}=\frac{1}{g}\tilde{Q}_{top}, \eeq where the first-class
topological charge $\tilde{Q}_{top}$ is given by \beq
\tilde{Q}_{top}=\frac{1}{4\pi}\int_{\tilde{S}^{2}_{(int)}}
d\tilde{A}_{a}^{(int)}\tilde{\bar{\phi}}^{a}, \eeq where
$\tilde{A}^{(int)}$ is the surface of a unit 2-sphere
$\tilde{S}^{2}_{(int)}$.  Here one notes that $\tilde{Q}_{top}$
yields the winding number in the map:
$S^{2}_{(phy)}\rightarrow\tilde{S}^{2}_{(int)}$, where
$S^{2}_{(phy)}$ and $\tilde{S}^{2}_{(int)}$ are the 2-sphere
compactified at infinity in the physical coordinate space and the
other 2-sphere of unit radius in the extended internal space of
$\tilde{\bar{\phi}}^{a}$ satisfying the first-class constraint
(\ref{constbart}), respectively, associated with the homotopy
group $\pi_{2}(\tilde{S}^{2})=Z$.  In the second-class
configuration, the topological charge $Q_{top}$ is described by
the winding number in the map: $S^{2}_{(phy)}\rightarrow
S^{2}_{(int)}$, where $S^{2}_{(int)}$ is the 2-sphere of unit
radius in the internal space of $\bar{\phi}^{a}$ with
$\bar{\phi}^{a}\bar{\phi}^{a}-1\approx 0$. The static conserved
energy $E$ of the 't Hooft-Polyakov monopole in the first-class
configuration, corresponding to the static limit of the
first-class Hamiltonian $\tilde{H}$ in (\ref{hct}) with
$\tilde{\lambda}=0$, is now given in terms of the
$\tilde{Q}_{top}$ \beq E=\int
d^{3}x\frac{1}{4}\left(\tilde{G}_{ij}^{a}
-\epsilon_{ijk}\tilde{D}^{k}\tilde{\phi}^{a}\right)^{2}+\frac{4\pi
F}{g}\tilde{Q}_{top},\label{bl} \eeq where the first-class
variables $\tilde{G}_{ij}^{a}$ and $\tilde{D}_{k}\tilde{\phi}^{a}$
are obtainable from Appendix A. For a given
$\tilde{Q}_{top}$-sector, the static energy $E$ has the
Bogomol'nyi lower bound $\frac{4\pi F}{g}\tilde{Q}_{top}$ when the
variables satisfy the condition \beq \tilde{G}_{ij}^{a}
=\epsilon_{ijk}\tilde{D}^{k}\tilde{\phi}^{a}.\label{bogo} \eeq

Now, in order to investigate the de Rham-type cohomology group
structure for the 't Hooft-Polyakov monopole system, we proceed
to implement the covariant Batalin-Fradkin-Vilkovisky
formalism~\cite{bfv}. We start by the construction of the
nilpotent BRST operator, by introducing two canonical sets
of ghost number $\eta=1$ field and ghost number $\eta=-1$ field
$({\cal C}^{i},\bar{{\cal P}}_{i})$, $({\cal P}^{i}, \bar{{\cal
C}}_{i})$ and the ghost number $\eta=0$ auxiliary fields $(N^{i},B_{i})$,
which satisfy the (anti)commutators, \beq \{{\cal C}^{i}(x),\bar{{\cal
P}}_{j}(y)\}=\{{\cal P}^{i}(x),\bar{{\cal
C}}_{j}(y)\}=\{N^{i}(x),B_{j}(y)\}=\delta_{j}^{i}\delta^{3}(x-y)~~~(i=1,2).
\eeq Here the super-Poisson bracket is defined as \beq
\{A,B\}=\frac{\delta A}{\delta q}|_{r}\frac{\delta B}{\delta
p}|_{l}-(-1)^{\eta_{A}\eta_{B}}\frac{\delta B}{\delta q}|_{r}\frac{\delta A} {%
\delta p}|_{l} \eeq where $\eta_{A}$ is the ghost number in $A$
and the subscript $r$ and $l$ denote right and left derivatives,
respectively.

The BRST operator for our constraint algebra is then simply given by
\beq
Q_{BRST} = \int {\rm d}^{3}x~({\cal C}^{i}\tilde{\Omega}_{i}+{\cal
P}^{i}B_{i}). \label{brstq} \eeq We choose the unitary gauge with
\beq\chi^{1}=\Omega_{1},~~~ \chi^{2}=\Omega_{2}\eeq by selecting
the gauge fixing functional \beq \Psi = \int {\rm
d}^{3}x~(\bar{{\cal C}}_{i}\chi^{i}+\bar{{\cal P}}_{i}N^{i}). \eeq
One can now readily see that $Q_{B}$ is nilpotent \beq
Q_{BRST}^{2}=\{Q_{BRST},Q_{BRST}\}=0, \eeq and $Q_{BRST}$ is the
generator of the infinitesimal BRST transformations \beq
\begin{array}{ll}
\delta_{Q_{BRST}}\phi^{a}=-{\cal C}^{2}\phi^{a},
&~~\delta_{Q_{BRST}}\pi^{a}=2{\cal C}^{1}\phi^{a}+{\cal
C}^{2}(\pi^{a}-2\phi^{a}\pi_{\theta}),\\
\delta_{Q_{BRST}}A_{\mu}^{a}=0, &~~\delta_{Q_{BRST}}\pi_{\mu}^{a}=0,\\
\delta_{Q_{BRET}}\theta={\cal C}^{2}\phi^{a}\phi^{a},
&~~\delta_{Q_{BRST}}\pi_{\theta}=2{\cal C}^{1},\\
\delta_{Q_{BRST}}{\cal C}^{i}=0,
&~~\delta_{Q_{BRST}}\bar{{\cal P}}_{i}=\tilde{\Omega}_{i},\\
\delta_{Q_{BRST}}{\cal P}^{i}=0,
&~~\delta_{Q_{BRST}}\bar{{\cal C}}_{i}=B_{i},\\
\delta_{Q_{BRST}}N^{i}=-{\cal P}^{i}, &~~\delta_{Q_{BRST}}B_{i}=0,\\
\delta_{Q_{BRST}}\lambda=0.
\end{array}
\label{brstgaugetrfm} \eeq

Furthermore, the first-class Hamiltonian $\tilde{H}$ in
(\ref{hct}) is $Q_{BRST}$-closed \beq
\delta_{Q_{BRST}}\tilde{H}=\{Q_{BRST},\tilde{H}\}=0,\eeq and \beq
\delta_{Q_{BRST}}\{Q_{BRST},\Psi\}=\{Q_{BRST},\{Q_{BRST},\Psi\}\}=0,
\label{gh} \eeq which follows from the nilpotency of the charge
$Q_{BRST}$. The gauge fixed BRST invariant Hamiltonian is now
given by
\bea H_{eff}&=&\tilde{H}^{\prime\prime}-\{Q_{BRST},\Psi\},\label{heffqbrst}\\
\tilde{H}^{\prime\prime}&=&\tilde{H}+\int
d^{3}x~\left(\pi_{\theta}\tilde{\Omega}_{2}-2{\cal C}^{1}\bar{\cal
P}_{2}\right), \label{heff} \eea with $\tilde{H}$ defined in
(\ref{hct}). In order to guarantee the BRST invariance of
$H_{eff}$, we have included in $H_{eff}$ of (\ref{heffqbrst}) the
$Q_{BRST}$-exact term, and in $\tilde{H}^{\prime \prime}$ of
(\ref{heff}) the term associated with $\pi_{\theta}$ in
$\tilde{H}^{\prime}$ in (\ref{hctp}) and the Faddeev-Popov ghost
term~\cite{faddev67}. In fact, the term $\{Q_{BRST},\Psi\}$ fixes
the particular unitary gauge corresponding to the fixed point
$(\theta=0, \pi_{\theta}=0)$ in the gauge degrees of freedom
associated with two dimensional manifold described by the internal
phase space coordinates $(\theta, \pi_{\theta})$, which physically
speaking are two canonically conjugate St\"uckelberg fields.

In general, by introducing the BRST operator \beq Q_{BRST}:
\omega_{p}\rightarrow \omega_{p+1}, \label{omegap} \eeq where
$\omega_{p}$ is a ghost number $p$-form with
$\eta_{\omega_{p}}=p$, we define the $p$-th de Rham-type
cohomology group $H^{p}(M,R)$ of the manifold $M$ and the field of
real number $R$ with the following quotient group \beq
H^{p}(M,R)=\frac{Z^{p}(M,R)}{B^{p}(M,R)}. \eeq Here $Z^{p}(M,R)$
are the collection of all $Q_{BRST}$-closed ghost number $p$-forms
$\omega_{p}$ for which $Q_{BRST}\omega_{p}=0$ and $B^{p}(M,R)$ are
the collection of all $Q_{BRST}$-exact ghost number $p$-forms
$\omega_{p}$ for which $\omega_{p}=Q_{BRST}\omega_{p-1}$.

For the case of the first-class Hamiltonian in (\ref{heff}), the
Hamiltonians $H_{eff}$ and $\tilde{H}^{\prime\prime}$ are readily
shown to be $Q_{BRST}$-closed as in the case of
$Q_{BRST}\Psi=\{Q_{BRST},\Psi\}$. With these ghost number
$0$-forms, we define the $Z^{0}(M,R)$ with $M$ being the monopole
Hilbert space and $R$ being the real field. Since $\Psi$ is the
ghost number $(-1)$-form and $Q_{BRST}\Psi$ is $Q_{BRST}$-exact
ghost number $0$-form, we also define the $B^{0}(M, R)$. Moreover,
the ghost number $0$-form $H_{eff}$ is deformed into the other
ghost number $0$-form $\tilde{H}^{\prime\prime}$. In other words,
$H_{eff}$ is homologous to $\tilde{H}^{\prime\prime}$ under the
BRST transformation $Q_{BRST}$,
$H_{eff}\sim\tilde{H}^{\prime\prime}$, since
$Q_{BRST}\Psi=\tilde{H}^{\prime\prime}-H_{eff}$.  With these
$Z^{0}(M,R)$ and $B^{0}(M,R)$, we define the $0$-th de Rham-type
cohomology group $H^{0}(M,R)$ for the 't Hooft-Polyakov monopole
system.

Finally, after the path integral algebra related to the
evaluation of the Legendre transformation of $H_{eff}$, we arrive
at the manifestly covariant BRST improved Lagrangian
\begin{equation}
L_{eff}= L_{0} + L_{WZ} + L_{ghost}
\label{lagfinal}
\end{equation}
where $L_{0}$ is given by (\ref{lag}) and
\begin{eqnarray}
L_{WZ}&=&\int d^{3}x~\left[
\frac{\theta}{\phi^{a}\phi^{a}}D_{\mu}\phi^{a}D^{\mu}\phi^{a}
-\lambda\theta(\phi^{a}\phi^{a}-F^{2}+\theta)-\frac{F^{2}}{2(\phi^{a}\phi^{a})^{2}}D_{\mu}
\theta D^{\mu}\theta\right],\nonumber\\
L_{ghost}&=&\int
d^{3}x~\left[-\frac{1}{2F^{2}}(\phi^{a}\phi^{a})^{2}
(B_{2}+2\bar{{\cal C}}_{2}{\cal C}^{2})^{2}
-\frac{1}{\phi^{a}\phi^{a}}D_{\mu}\theta D^{\mu}B_{2}
+D_{\mu}\bar{{\cal C}}_{2}D^{\mu}{\cal C}^{2}\right].
\label{lagwz}
\end{eqnarray}
The Lagrangian $L_{eff}$ in (\ref{lagfinal}) can be readily shown to
be a covariant form of the 't Hooft-Polyakov monopole Lagrangian in (\ref{lag}).
Here, we note that the canonical fields
$(\phi^{a},A_{\mu}^{a},\lambda)$ in $L_{eff}$ are unconstrained
ones and the  St\"uckelberg field $\theta$
becomes a nontrivial propagating field. The BRST gauge fixed
effective Lagrangian (\ref{lagfinal}) is readily shown to be
manifestly invariant under the following BRST transformations, \beq
\begin{array}{ll}
\delta_{\epsilon}\phi^{a}=\epsilon \phi^{a}{\cal C}^{2},
&~~\delta_{\epsilon}A_{\mu}^{a}=0,\\
\delta_{\epsilon}\lambda=0, &~~\delta_{\epsilon}\theta=-\epsilon
\phi^{a}\phi ^{a}{\cal C}^{2},\\
\delta_{\epsilon}\bar{{\cal C}}_{2}=-\epsilon B_{2},
&~~\delta_{\epsilon}{\cal C}^{2}=\delta_{\epsilon}B_{2}=0,\\
\end{array}
\eeq
where $\epsilon$ is an infinitesimal Grassmann valued
parameter.

\section{Conclusions}
\setcounter{equation}{0}
\renewcommand{\theequation}{\arabic{section}.\arabic{equation}}

In conclusion, we have found the closed algebra
of the quantum operators in the the 't Hooft-Polyakov monopole using
the Dirac brackets of the Dirac Hamiltonian formalism. Next, in the
first-class configuration, we have constructed the first-class monopole
charge in the U(1) subgroup of the SU(2) gauge group to discuss the
Bogomol'nyi bound defined on the extended internal 2-sphere.  We then
have included the $Q_{BRST}$-exact gauge fixing term and the
Faddeev-Popov ghost term in the first-class Hamiltonian to define
the BRST symmetries and the de Rham-type cohomology group for the
monopole system.

The cohomology group structure discussed in this paper is generic
property shared by the constrained physical systems, such as the
monopoles, Skyrmion model, CP(n) model, O(n) model, and so
on~\cite{hong02pr}. In these models, the fields $\phi^{a}$ satisfy
the geometric constraint of the typical form
$\phi^{a}\phi^{a}-F^{2}\approx 0$. Using the time evolution of
this constraint, we construct the second-class constraints of the
models, which can be converted into the first-class systems in the
Dirac Hamiltonian quantization by introducing the St\"uckelberg
fields. Next by including the ghost degrees of freedom in the
constrained models, we can discuss the BRST symmetries and also
define the de Rham-type cohomology group as in this 't
Hooft-Polyakov monopole system. In the future studies, it will be
interesting to discuss the generic features of the cocycles
involved in the constrained physical systems.

\acknowledgments The author would like to thank J. Lee, T.H. Lee,
A.J. Niemi and P. Oh for helpful discussions. This work was
supported by the Korea Research Foundation (MOEHRD), Grant No.
KRF-2006-331-C00071.

\appendix
\section{First-class canonical variables}
\setcounter{equation}{0}
\renewcommand{\theequation}{A.\arabic{equation}}

We construct the first-class canonical variables $\tilde{{\cal
F}}
=(\tilde{\phi}^{a},\tilde{\pi}^{a},\tilde{A}_{\mu}^{a},\tilde{\pi}_{\mu}^{a},
\tilde{\lambda},\tilde{\pi}_{\lambda})$, associated with the
original variables ${\cal
F}=(\phi^{a},\pi^{a},A_{\mu}^{a},\pi_{\mu}^{a},\lambda,\pi_{\lambda}
)$, in the extended phase space. These variables are obtained as a
power series in the St\"uckelberg fields $(\theta,\pi_{\theta})$,
by demanding that they are in strong involution with the
first-class constraints (\ref{1stconst}), \beq
\{\tilde{\Omega}_{i}, \tilde{{\cal F}}\}=0.\eeq After some algebra
similar to the case of the first-class Hamiltonian, we obtain for
the first-class canonical variables \bea
\tilde{\phi}^{a}&=&\phi^{a}\left(\frac{\phi^{c}\phi^{c}
+2\theta}{\phi^{c}\phi^{c}}\right)^{1/2},\nn\\
\tilde{\pi}^{a}&=&\left(\pi^{a}-\phi^{a}\pi_{\theta}\right)
\left(\frac{\phi^{c}\phi^{c}}{\phi^{c}\phi^{c}+2\theta}\right)^{1/2},\nn\\
\tilde{A}_{\mu}^{a}&=&A_{\mu}^{a},~~~\tilde{\pi}_{\mu}^{a}=\pi_{\mu}^{a},\nn\\
\tilde{\lambda}&=&\lambda,~~~~~\tilde{\pi}_{\lambda}=\pi_{\lambda}.
\label{pitilde} \eea

Next, we find for the Hamiltonian in (\ref{hct}) \beq
\tilde{H}=\int
d^{3}x\left[\frac{1}{2}\tilde{\pi}^{a}\tilde{\pi}^{a}
+\frac{1}{2}\tilde{\pi}_{i}^{a}\tilde{\pi}_{i}^{a}
+\frac{1}{2}\tilde{D}_{i}\tilde{\phi}^{a}\tilde{D}_{i}\tilde{\phi}^{a}
+\frac{1}{4}\tilde{G}_{ij}^{a}\tilde{G}_{ij}^{a}+\frac{1}{4}
\tilde{\lambda}(\tilde{\phi}^{a}\tilde{\phi}^{a}-F^{2})^{2}\right],
\label{htilde} \eeq where the first-class variables are given
above and \bea
\tilde{D}_{i}\tilde{\phi}^{a}&=&D_{i}\phi^{a}\left(\frac{\phi^{c}\phi^{c}
+2\theta}{\phi^{c}\phi^{c}}\right)^{1/2},\nn\\
\tilde{G}_{ij}^{a}&=&G_{ij}^{a}. \eea

\end{document}